\documentclass{article}
\usepackage{graphicx}
\usepackage{typearea}
\typearea{20}

\providecommand{\keywords}[1]
{
  \small	
  \textbf{\textit{Keywords---}} #1
}

\title{Fairness in Proof of Team Sprint (PoTS): \\
Evaluating Reward Distribution Across Performance Levels}
\author{Naoki Yonezawa \thanks{\texttt{n.yonezawa@thu.ac.jp}. Faculty of Humanities and Social Sciences, Teikyo Heisei University, Japan.}}
\date{}

\begin{document}

\maketitle

\begin{abstract}
Blockchain consensus mechanisms must balance security, decentralization, and efficiency while ensuring fair participation. Proof of Team Sprint (PoTS) is a cooperative consensus mechanism designed to address the energy inefficiencies and centralization tendencies of traditional Proof of Work (PoW). Unlike PoW, where rewards disproportionately favor high-performance nodes, PoTS encourages collaboration by forming teams and distributing rewards more equitably among participants. In this study, we evaluate the fairness properties of PoTS by analyzing reward distribution under varying computational power distributions. Through extensive simulations, we compare equal-share allocation and proportional reward allocation, highlighting their impact on decentralization and participation. Our results demonstrate that PoTS significantly reduces reward disparity between high-performance and low-performance nodes, fostering a more inclusive ecosystem. Additionally, we observe that as team sizes increase, the influence of individual computational power is mitigated, allowing lower-performance nodes to contribute meaningfully. Moreover, our findings reveal that the marginal benefit of investing in extremely high-performance hardware diminishes, which discourages centralization and aligns incentives toward sustainable participation. We also discuss the economic implications of PoTS, particularly its potential to reshape blockchain mining strategies by balancing fairness with computational efficiency. These insights contribute to the broader discussion on blockchain fairness and provide a foundation for further research into cooperative consensus mechanisms.
\end{abstract}


\section{Introduction}

\subsection{Background and Motivation}
Blockchain technology has revolutionized decentralized systems, enabling trustless peer-to-peer transactions and removing reliance on central authorities. Bitcoin \cite{nakamoto2008peer}, as the first successful implementation of blockchain, introduced the Proof-of-Work (PoW) consensus mechanism. While PoW has ensured network security and decentralization, it has faced significant criticisms regarding its fairness and energy inefficiency. The competitive nature of PoW leads to an uneven distribution of rewards, where participants with superior computational resources receive disproportionately higher incentives. Consequently, mining power is concentrated among a small number of entities, raising concerns about centralization and accessibility. Furthermore, the extreme energy consumption required for PoW computations has been widely debated as a critical environmental challenge.

To address these limitations, alternative consensus mechanisms have emerged, including Proof-of-Stake (PoS) and cooperative models such as Proof of Team Sprint (PoTS) \cite{yonezawa2024pots}. PoTS represents a shift from adversarial mining competition to a collaborative framework where participants form teams to contribute to consensus. Unlike PoW, which disproportionately favors high-performance nodes, PoTS distributes rewards based on team performance, ensuring that even low-performance participants receive incentives. This design not only promotes decentralization by reducing wealth concentration but also leads to a more energy-efficient blockchain network.

\begin{figure}[t]
\begin{center}
\includegraphics[scale=0.5]{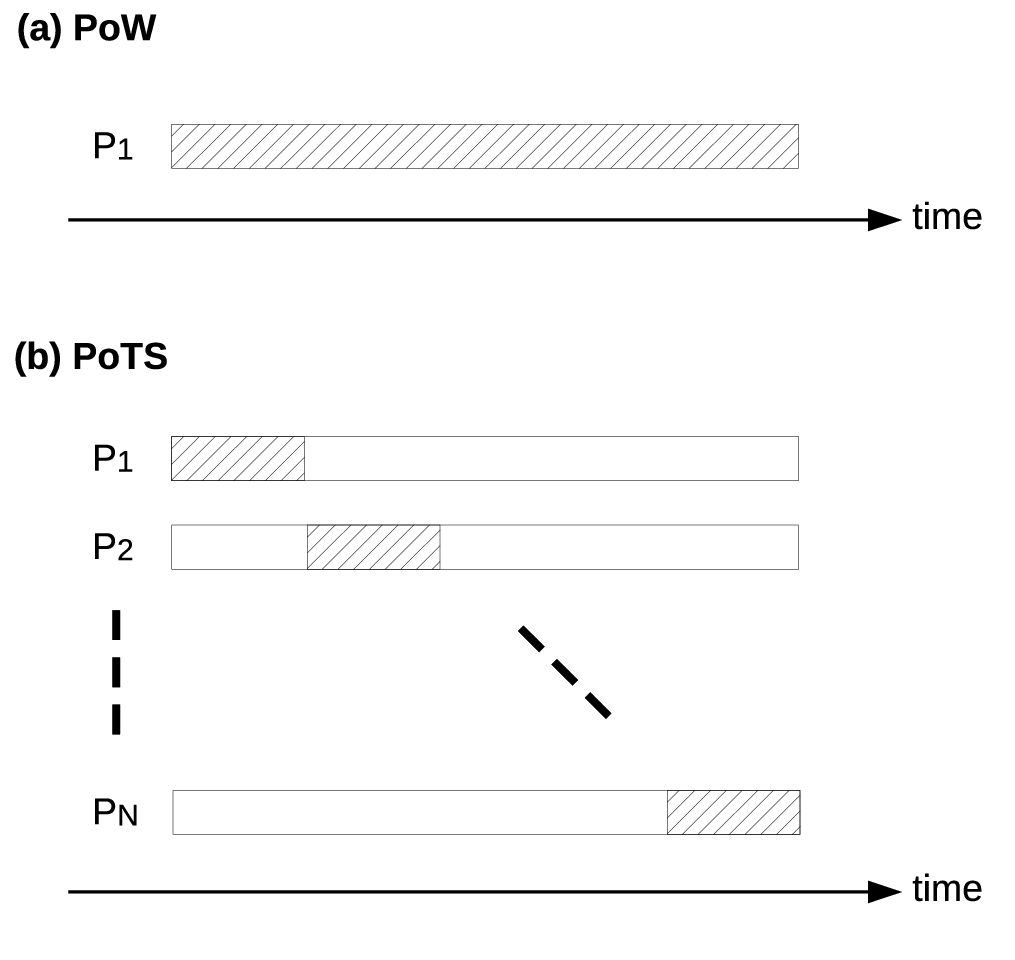}
\caption{Comparison of Block Generation Process in PoW and PoTS. Adapted from~\cite{yonezawa2024pots}.}
\label{fig:pow-pots-time-diagram}
\end{center}
\end{figure}

The fundamental difference between PoW and PoTS in the block generation process is illustrated in Figure~\ref{fig:pow-pots-time-diagram}. While PoW requires a single participant to complete an entire computational task before receiving rewards, PoTS distributes computational efforts among multiple participants, who work collaboratively in a sequential manner. This cooperative approach reduces computational redundancy and increases overall network efficiency.

Fairness is a fundamental principle in blockchain consensus mechanisms, influencing economic inclusivity, system security, and long-term network sustainability. Traditional models often exacerbate economic disparities by rewarding only the most resource-intensive participants, discouraging engagement from those with limited computational power. By contrast, PoTS aims to create an environment where diverse participants can meaningfully contribute to the consensus process. Analyzing the fairness properties of PoTS is therefore essential in assessing its potential as a viable alternative to existing mechanisms, particularly in ensuring equitable reward distribution and sustainable participation.

\subsection{Research Contributions}
This study aims to provide an in-depth evaluation of the fairness properties of PoTS, particularly in terms of reward distribution and computational efficiency. Our primary contributions are as follows:
\begin{itemize}
    \item We conduct extensive simulations to analyze and compare equal-share allocation and proportional reward allocation schemes in PoTS, quantifying their effects on reward fairness.
    \item We demonstrate that PoTS provides greater reward opportunities for low-performance participants, in contrast to PoW, where mining rewards are monopolized by high-performance entities.
    \item We examine the phenomenon of diminishing returns for high-performance nodes, highlighting its implications for discouraging excessive centralization and promoting balanced participation.
    \item We explore the broader implications of PoTS fairness on blockchain ecosystems, considering its potential impact on decentralization, energy efficiency, and long-term network stability.
\end{itemize}
Understanding these aspects is crucial for determining the viability of PoTS as a fair and sustainable consensus mechanism. By fostering a more inclusive and cooperative blockchain environment, PoTS has the potential to redefine the incentive structures that drive decentralized networks.

The remainder of this paper is structured as follows. Section \ref{sec:related_work} reviews prior research on fairness in blockchain consensus mechanisms. Section \ref{sec:methodology} describes our simulation framework and experimental setup. Section \ref{sec:results} presents our findings on reward distribution and efficiency. Section \ref{sec:discussion} discusses the implications of these results. Finally, Section \ref{sec:conclusion} concludes the paper and outlines directions for future research.

\section{Related Work} \label{sec:related_work}

The study of fairness, reward distribution, and incentive mechanisms in blockchain has been extensively researched. This section discusses prior work on blockchain fairness, cooperative reward-sharing models, and economic perspectives on incentive structures.

\subsection{Fairness in Blockchain Consensus Mechanisms}
Fairness in blockchain ecosystems has been a persistent concern, especially in Proof-of-Work (PoW) and Proof-of-Stake (PoS) protocols. 

Afonso \cite{afonso2024fairness} highlights systemic fairness challenges in blockchain consensus mechanisms, emphasizing how certain protocols inherently favor well-resourced participants. Similarly, Huang \textit{et al.} \cite{huang2021rich} provides a detailed fairness analysis, demonstrating how existing reward structures tend to reinforce economic inequalities among participants.

To address these concerns, alternative blockchain designs have been proposed. FruitChains introduces a fairer approach to reward distribution by ensuring that miners receive rewards proportional to their computational effort, mitigating selfish mining strategies \cite{pass2017fruitchains}. In committee-based blockchain systems, fairness considerations have also been explored, ensuring that consensus decisions are not dominated by a small subset of participants \cite{amoussou2019fairness}. Moreover, order-fairness in Byzantine consensus mechanisms has been studied as a means to prevent manipulations by high-resource actors, reinforcing the need for fair execution in transaction ordering \cite{kelkar2020order}.

Alamer \textit{et al.} \cite{alamer2024proof} propose Proof of Fairness, a dynamic and secure consensus protocol designed to enhance fairness and security in blockchain operations. Zheng \textit{et al.} \cite{zheng2025justitia} introduce Justitia, an incentive mechanism aimed at ensuring fairness in cross-shard transactions, mitigating potential imbalances in distributed environments. Furthermore, Orda \textit{et al.} \cite{orda2021enforcing} investigate enforcement strategies for fairness in blockchain transaction ordering, addressing concerns about manipulation by high-resource participants.

Yu \textit{et al.} \cite{yu2018survey} provide a comprehensive survey on consensus and incentive mechanisms derived from P2P networks, highlighting how fairness considerations affect blockchain system design. G\"und\"uz \textit{et al.} \cite{gunduz2023proof} propose the Proof of Optimum (PoO) model, which balances fairness and efficiency, ensuring that participation is not disproportionately influenced by computational power. Abraham \textit{et al.} \cite{abraham2016solidus} introduce Solidus, an incentive-compatible cryptocurrency based on permissionless Byzantine consensus, addressing concerns related to fair participation and resistance to manipulation. Yuming \cite{yuming2023fairness} explore fairness and security aspects in blockchain incentive mechanisms, emphasizing the need for equitable participation.

\subsection{Reward Distribution in Cooperative Mechanisms}
To mitigate centralization risks in mining, several decentralized reward-sharing models have been explored. Gudmundsson \textit{et al.} \cite{gudmundsson2024blockchain} investigate how decentralized mining pools can achieve fairer reward distribution without relying on centralized authority. Additionally, Zhao \textit{et al.} \cite{zhao2022bayesian} propose a Bayesian-Nash-compatible mechanism for transaction fee allocation, ensuring incentive compatibility in blockchain environments.

Ethereum 2.0's transition to Proof-of-Stake (PoS) has sparked research on reward dynamics and decentralization. The work in Yan \textit{et al.} \cite{yan2024analyzing} examines staking rewards and their implications for decentralization, providing insights into how PoS-based networks can maintain fairness. Additionally, the Colordag blockchain proposes an incentive-compatible structure for fair reward allocation, minimizing economic disparity \cite{abraham2023colordag}.

Chen \textit{et al.} \cite{chen2024fairreward} present FairReward, which applies equity theory to achieve fair reward distribution in blockchain-based federated learning. Li \textit{et al.} \cite{li2023reward} analyze reward distribution trade-offs in Proof-of-Stake protocols, balancing inclusion and fairness. Br\"unjes \textit{et al.} \cite{brunjes2020reward} investigate reward sharing schemes for stake pools, optimizing incentive models for long-term sustainability.

He \textit{et al.} \cite{he2018blockchain} develop a blockchain-based truthful incentive mechanism for distributed P2P applications, ensuring that incentives align with truthful reporting. Similarly, He \textit{et al.} \cite{he2022blockchain} propose a content delivery network leveraging blockchain-based monetary incentives, guaranteeing fairness in data dissemination. The BIMEE model, which considers the endowment effect in blockchain-based incentives, adjusts reward mechanisms to reflect psychological factors in decision-making \cite{madupalli2023bimee}.

\subsection{Economic and Game-Theoretic Perspectives}
Game-theoretic models have played a crucial role in analyzing blockchain incentives. The study in Li \textit{et al.} \cite{li2023game} applies game theory to non-cryptocurrency blockchain incentive design, demonstrating how strategic incentives can align with decentralized consensus. Additionally, Han \textit{et al.} \cite{han2022can} explore the broader relationship between blockchain incentive mechanisms and economic sustainability.

The impact of economic disparities on user engagement has also been examined in the context of crypto-reward fairness. Research in Yang \textit{et al.} \cite{yang2024whales} investigates how reward imbalances affect participation in blockchain-based economies. Blockchain-based decentralized infrastructures have also been analyzed for their boundary conditions, revealing limitations in existing economic models \cite{pereira2019blockchain}. Furthermore, smart contract-based incentive mechanisms have been proposed as a solution to improving fairness in decentralized networks \cite{xuan2020incentive}.

Hu \textit{et al.} \cite{hu2021hybrid} explore hybrid blockchain designs for IoT applications, analyzing energy efficiency and reward mechanisms. Mancino \textit{et al.} \cite{mancino2025striking} study content quality and reward dynamics in blockchain-based social media, assessing the trade-offs between engagement incentives and fairness. Finally, Xiong \textit{et al.} \cite{xiong2022research} review recent advancements in blockchain consensus algorithms, with a focus on economic implications and incentive mechanisms. Zhu \textit{et al.} \cite{zhu2020improved} propose an improved Proof-of-Trust consensus algorithm that enhances credibility in crowdsourcing blockchain services, integrating trust-based metrics into the consensus process. Liu \textit{et al.} \cite{liu2022incentive} present an incentive mechanism for sustainable blockchain storage, ensuring long-term participation remains economically viable.

\section{Methodology} \label{sec:methodology}

\subsection{Simulation Framework}
This section describes the framework used for our simulation study. The simulation models a competitive environment where participants with different computational capabilities form teams and compete in mining a fictitious cryptocurrency, SprintCoin, blocks. The core elements of the framework include the team size, the distribution of computational performance among participants, and the reward allocation mechanisms.

The simulation consists of $n$ participants, each assigned a computational performance level according to a predefined distribution. In this study, $n$ is fixed at 1600. Participants are grouped into teams of size $N$, where $N$ takes values from ${1, 2, 4, 8, 16, 32, 64}$. Each team attempts to complete a mining task that represents a sprint consisting of a predefined amount of work, denoted as 600 seconds of base computational time. The individual workload for each team member is approximately $600/N$ seconds but varies within a range determined by a uniformly sampled factor in $[0.8,1.2]$ of this base value to introduce randomness in task completion time. The first team to complete its assigned work wins the round and receives a reward of 10 units of SprintCoin. This process repeats for 1000 rounds, with teams being reshuffled after each round to ensure fairness and variation in results.

In this simulation, we assume an idealized environment in which there is no network latency or communication overhead among participants. Furthermore, all participants are assumed to behave reliably and to complete their assigned computational tasks without interruption or dropout. These assumptions are made to isolate the effects of performance distribution and reward allocation schemes, allowing us to focus on the fairness and efficiency of the system under controlled and deterministic conditions.

Two reward allocation strategies are evaluated: equal-share allocation and proportional reward allocation. In the equal-share allocation scheme, all members of the winning team receive an equal portion of the reward, regardless of their computational contribution. In the proportional reward allocation scheme, the reward is distributed based on each member’s relative contribution to the team’s total computational power. The impact of these reward mechanisms is analyzed across different performance distributions.

\subsection{Performance Distributions Analyzed}

To evaluate the impact of computational power heterogeneity, we consider multiple performance distributions. Each distribution defines the relative performance of participants in the simulation.

In general, a performance distribution can be expressed as:
\[
\{ p_1 : n_1, p_2 : n_2, \dots, p_k : n_k \}
\]
where \( p_i \) represents the performance level of a group of participants, and \( n_i \) denotes the number of participants with that performance level. The total number of participants satisfies:
\[
\sum_{i=1}^{k} n_i = n.
\]

The distributions analyzed in this study are:

\begin{itemize}
    \item $\{1:800,r:800\}$ \quad (Scenarios where two performance classes exist, with one group having performance level 1 and the other having performance level $r$.)
    \item $\{1:1,r:1599\}$ \quad (Scenarios where a single low-performance participant competes against $1599$ high-performance participants.)
    \item $\{1:1599,r:1\}$ \quad (Scenarios where a single high-performance participant competes against $1599$ low-performance participants.)
    \item $\{1:800,2:400,3:200,4:100,5:50,6:30,7:10,8:5,9:3,10:2\}$ \quad (A multi-layered performance distribution representing a more complex computational landscape.)
\end{itemize}

Each scenario provides insight into different aspects of fairness and reward distribution. The impact of computational disparities is analyzed by varying $r$ among ${2, 5, 10, 100}$ in the first three cases. However, due to space constraints, the cases of $r = 5$ and $r = 10$ are omitted.

\subsection{Simulation Execution}

The simulation proceeds as follows:

\begin{enumerate}
    \item $n$ participants are initialized.
    \item Each participant is assigned a performance level based on a predefined performance distribution.
    \item Participants are randomly assigned to teams of size $N$.
    \item Each team attempts to mine SprintCoin blocks within an allotted time window of 600 seconds (base time).
    \item Individual work assignments are distributed among team members, with each node handling approximately $600/N$ seconds of work. 
          The actual workload is determined by a uniformly distributed random scaling factor within the range $[0.8, 1.2]$ of the assigned task duration.
          Additionally, the processing time is further shortened based on each participant's computational performance.
    \item The first team to complete the required number of blocks (equal to the team size $N$) wins and receives a reward of 10 SprintCoin units.
    \item The winning team's reward is either distributed equally or according to proportional reward allocation.
    \item Teams are reshuffled after each round, and the process repeats for 1000 rounds.
    \item After 1000 rounds, participant rewards are calculated and stored, and the average reward and efficiency for each performance level are computed.
    \item Each simulation configuration is executed 100 times to ensure statistical significance, and final results are obtained by averaging across runs.
\end{enumerate}

\subsection{Evaluation Metrics}
To assess the effectiveness of different reward allocation strategies, the following metrics are used:

\begin{itemize}
    \item \textbf{Average Reward}: The mean amount of SprintCoin earned by participants at a given performance level across all rounds.
    \item \textbf{Efficiency}: Defined as the average reward per unit of computational performance. This metric assesses the fairness of reward distribution by evaluating how well rewards are balanced relative to performance levels.
\end{itemize}

By analyzing these metrics, we aim to understand whether lower-performance participants have a viable path to earning rewards under different reward allocation schemes and whether investing in extremely high-performance machines results in proportionally higher rewards.

\section{Results and Discussion}\label{sec:results}

\subsection{Reward Distribution Across Performance Levels}
Our simulations explored various performance distribution scenarios to evaluate reward distribution patterns under PoTS.

\begin{figure}[htbp]
\centering
\includegraphics[width=0.7\linewidth]{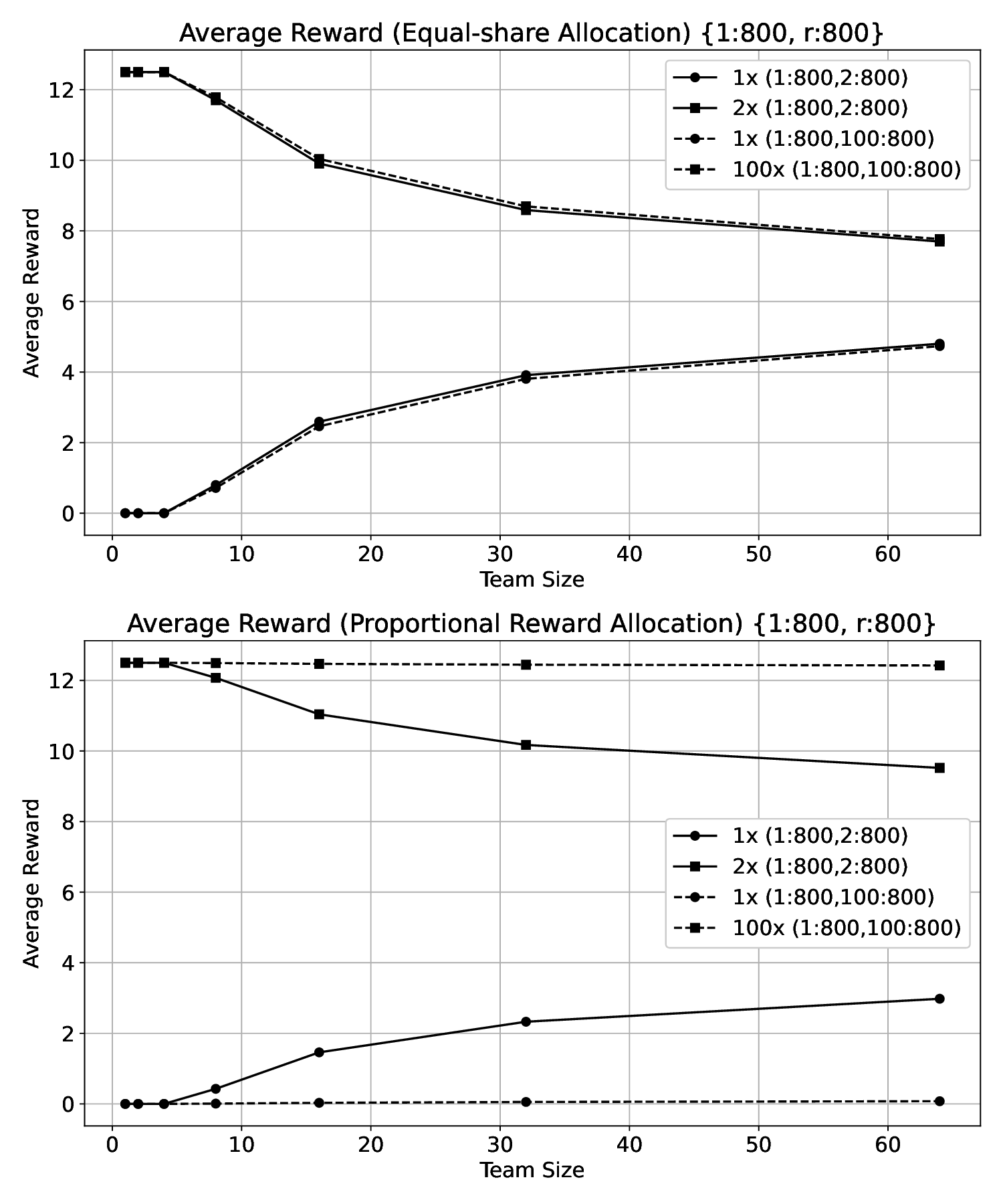}
\caption{Average reward distribution for equal-share allocation and proportional reward allocation schemes in \{1:800, $r$:800\}.}
\label{fig:average_reward_800_800}
\end{figure}

\paragraph{Performance Distribution \{1:800, $r$:800\} ($r = 2, 100$)}
Figure~\ref{fig:average_reward_800_800} illustrates the reward distribution trends under equal-share allocation and proportional reward allocation for performance distribution \{1:800, $r$:800\}.

Under equal-share allocation, the average rewards for low-performance and high-performance nodes remain nearly identical regardless of $r$. However, increasing the team size causes a decline in high-performance node rewards while increasing the rewards for low-performance nodes. This trend suggests that as team sizes grow, the team's overall ability is averaged, making it easier for lower-performance groups to succeed.

Conversely, under proportional reward allocation, team size also helps balance the rewards between low- and high-performance nodes, but to a lesser extent than in equal-share allocation. When $r = 2$, the gap between the two groups narrows with increasing team size, whereas when $r = 100$, the gap remains significantly large. This outcome highlights that performance differences persist under proportional reward allocation but are somewhat mitigated as team sizes increase.

\paragraph{Extreme Performance Distributions}We further analyze extreme cases to highlight PoTS's capability to provide inclusive incentives even in highly imbalanced scenarios.

\begin{figure}[htbp]
\centering
\includegraphics[width=0.7\linewidth]{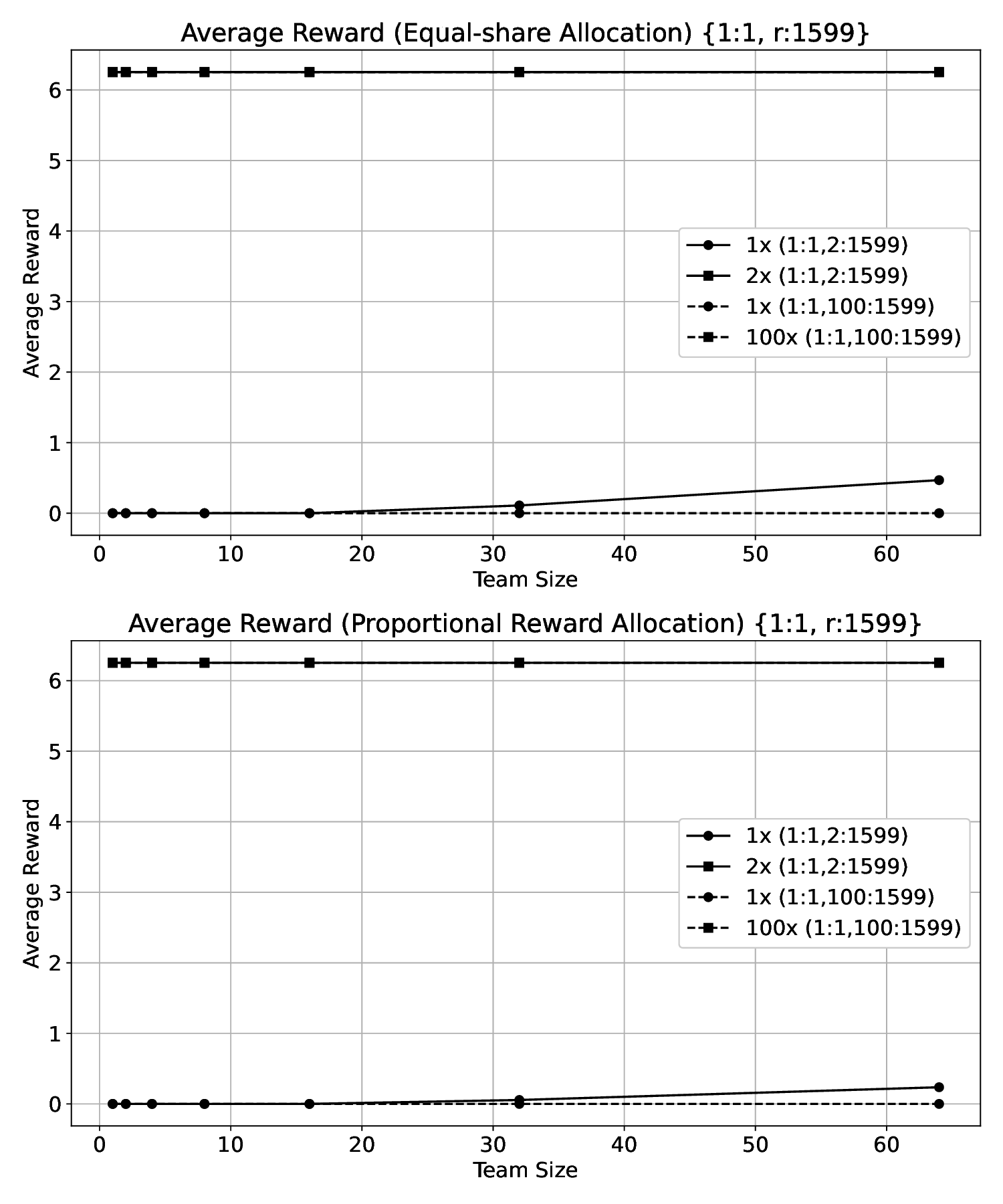}
\caption{Average reward distribution for \{1:1, $r$:1599\}.}
\label{fig:average_reward_1_1599}
\end{figure}

For the distribution \{1:1, $r$:1599\}, Figure~\ref{fig:average_reward_1_1599} shows that when the performance ratio is moderate ($r = 2$), low-performance node begins earning rewards at team sizes of 32 or larger. However, under extreme conditions ($r = 100$), low-performance node receives no rewards, emphasizing the challenges posed by extreme disparities. Notably, there is minimal difference between equal-share allocation and proportional reward allocation, reinforcing the notion that PoTS enables reward distribution even under such extreme settings.

\begin{figure}[htbp]
\centering
\includegraphics[width=0.7\linewidth]{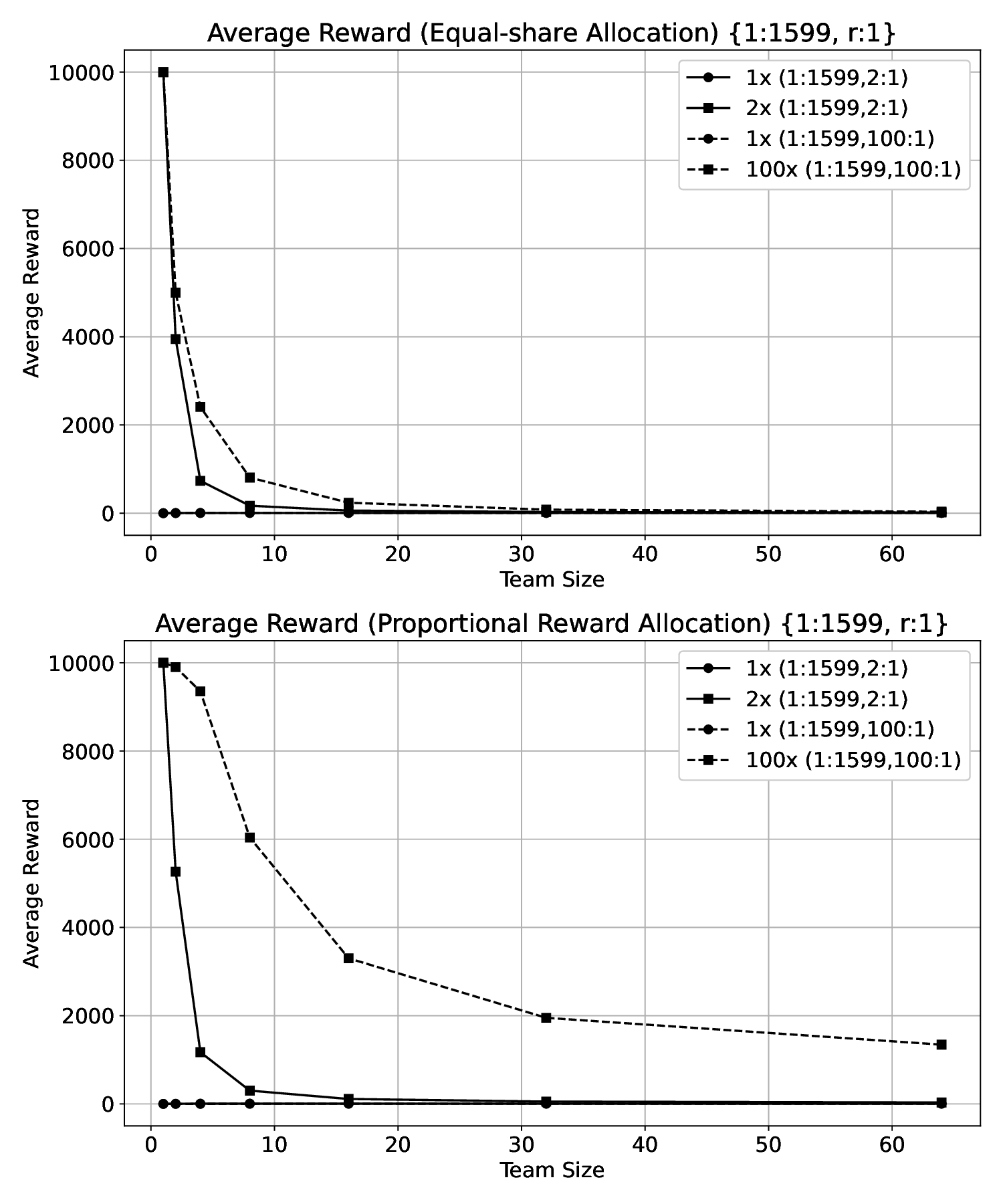}
\caption{Average reward distribution for \{1:1599, $r$:1\}.}
\label{fig:average_reward_1599_1}
\end{figure}

For the distribution \{1:1599, $r$:1\}, Figure~\ref{fig:average_reward_1599_1} illustrates a striking shift: when $N=1$, rewards are monopolized by the high-performance node, mimicking PoW. However, as team sizes increase, the dominance of high-performance node quickly diminishes, leading to a fairer distribution of rewards.

Under equal-share allocation, for $r = 2$ and $N=64$, low-performance node achieves an average reward of 6.24 coins compared to 16.50 coins for high-performance node. When $r = 100$, low-performance node's reward remains stable at 6.23 coins, while high-performance nodes gain 34.22 coins.

Under proportional-share allocation, for $r = 2$ and $N=64$, low-performance node receives 6.23 coins while high-performance nodes receive 32.49 coins. However, when $r = 100$, high-performance nodes achieve drastically higher rewards (1,343.50 coins), highlighting the large disparity.

\begin{figure}[htbp]
\centering
\includegraphics[width=0.7\linewidth]{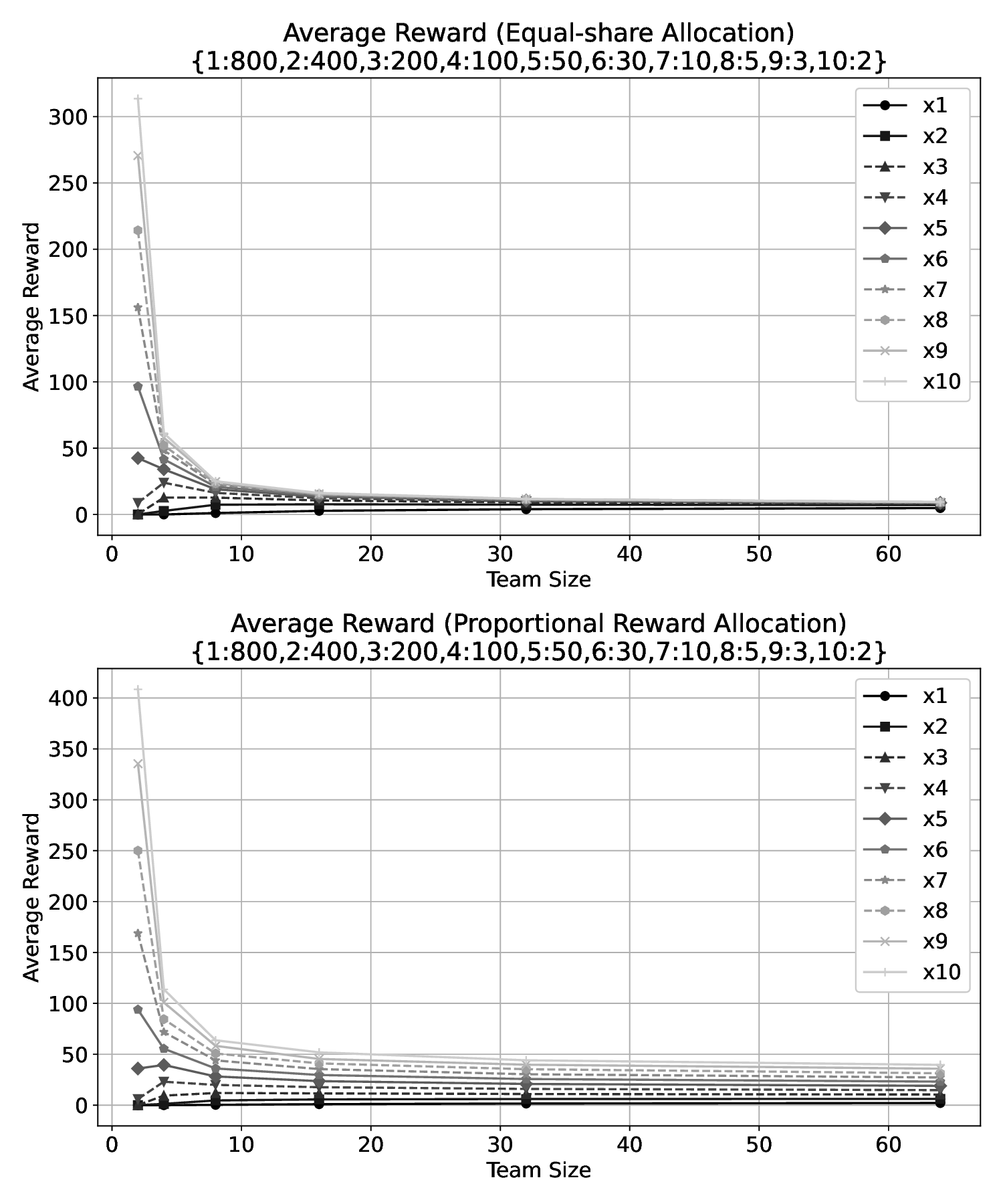}
\caption{Average reward distribution across multiple performance levels.}
\label{fig:average_reward_combined_10layers}
\end{figure}

\paragraph{Multi-Layered Performance Distribution}
A more realistic performance distribution is presented in Figure~\ref{fig:average_reward_combined_10layers}, where computational power is distributed across multiple layers: \{1:800,2:400,3:200,4:100,5:50,6:30,7:10,8:5,9:3,10:2\}. 

In this scenario, when team size $M=1$, the highest-performance group dominates, leading to extreme values (omitted from the graph). However, as team sizes increase, the reward disparities between different performance levels decrease under both equal-share allocation and proportional reward allocation. This finding suggests that PoTS can effectively balance reward distribution in a heterogeneous computing environment, making it a viable consensus mechanism for real-world applications.

\subsection{Efficiency Analysis}
Efficiency is defined as the average reward obtained per unit of computational performance. This metric provides insight into the cost-effectiveness of participating nodes under different allocation strategies.

\begin{figure}[htbp]
\centering
\includegraphics[width=0.7\linewidth]{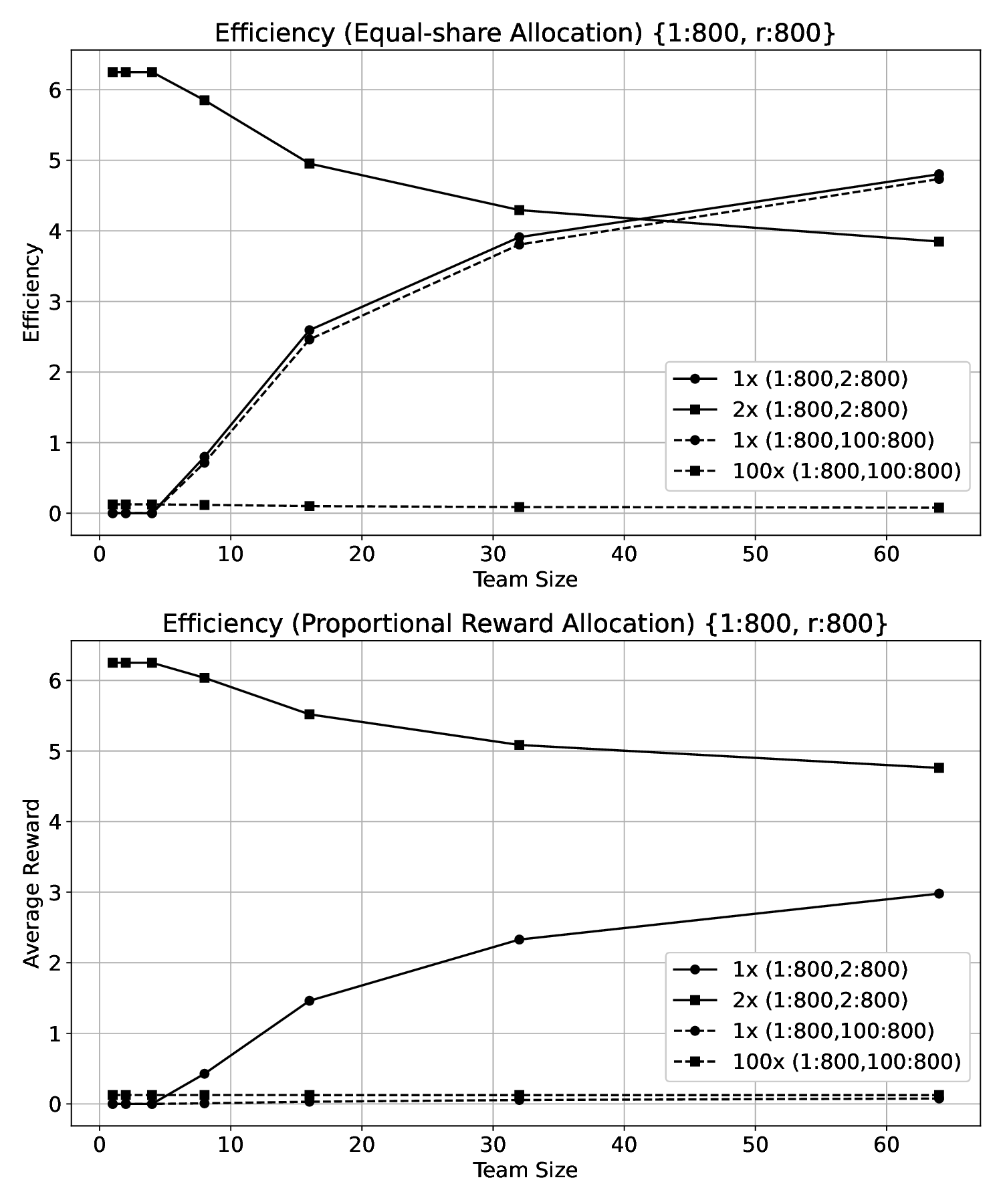}
\caption{Efficiency comparison for equal-share and proportional-share schemes in {1:800, $r$:800}.}
\label{fig:efficiency_800_800}
\end{figure}

\paragraph{Performance Distribution {1:800, $r$:800} ($r = 2, 100$)}
Figure~\ref{fig:efficiency_800_800} presents the efficiency trends for this performance distribution.

In the equal-share allocation scheme, the efficiency values reveal a significant trend: at $M=64$, the efficiency of low-performance nodes surpasses that of high-performance nodes. This means that lower-performance nodes achieve a better cost-performance ratio. For high-performance nodes with a $100\times$ performance advantage, participation becomes highly inefficient, making it an unfavorable investment.

\paragraph{Extreme Performance Distributions}
We further analyze extreme cases to highlight the efficiency trends in highly imbalanced scenarios.

\begin{figure}[htbp]
\centering
\includegraphics[width=0.7\linewidth]{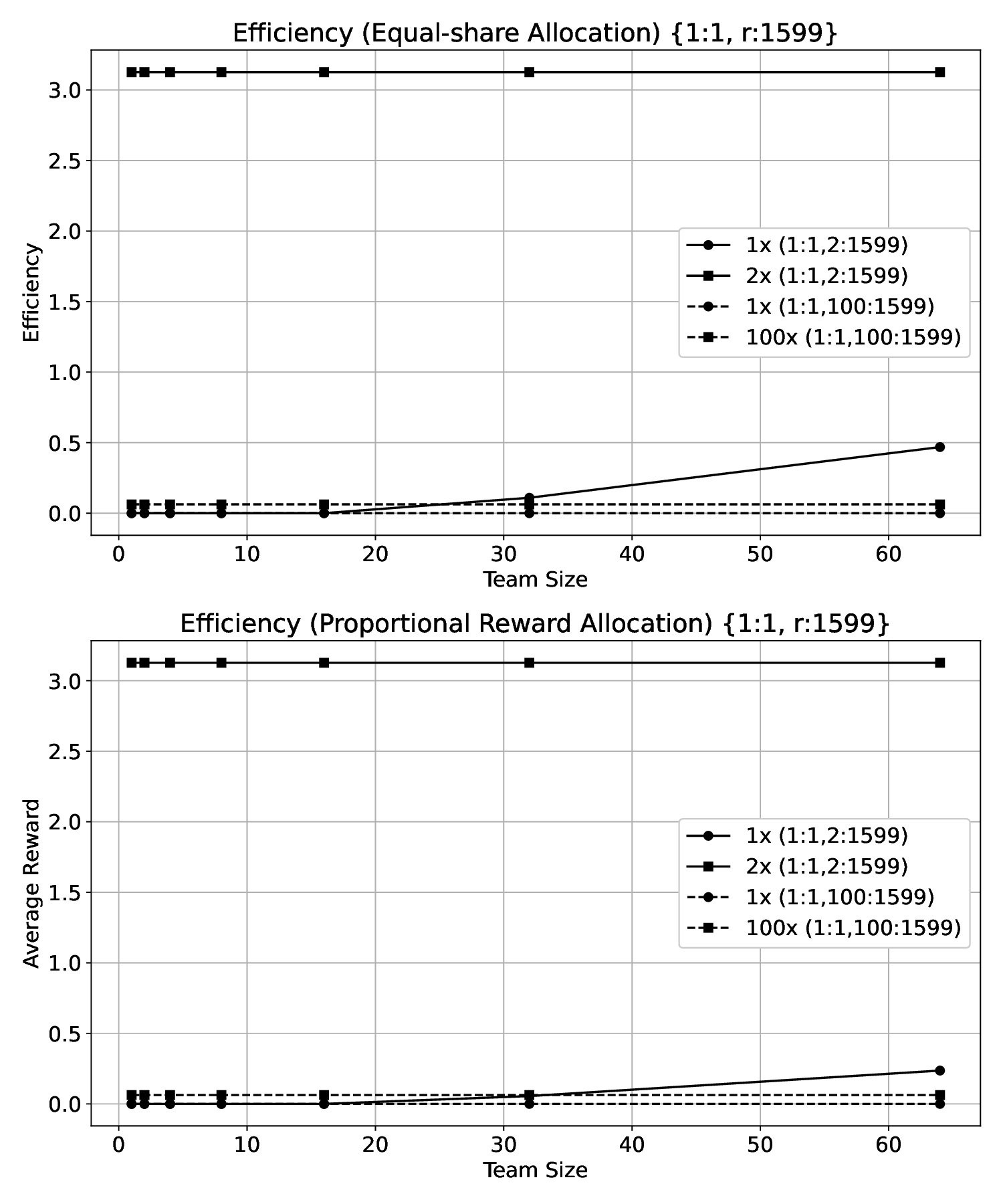}
\caption{Efficiency comparison for {1:1, $r$:1599}.}
\label{fig:efficiency_1_1599}
\end{figure}

For the distribution \{1:1, $r$:1599\}, Figure~\ref{fig:efficiency_1_1599} shows that the low-performance node has a minimal effect, and efficiency inversion does not occur. Again, for the $100\times$ performance nodes, investment efficiency remains poor.

\begin{figure}[htbp]
\centering
\includegraphics[width=0.7\linewidth]{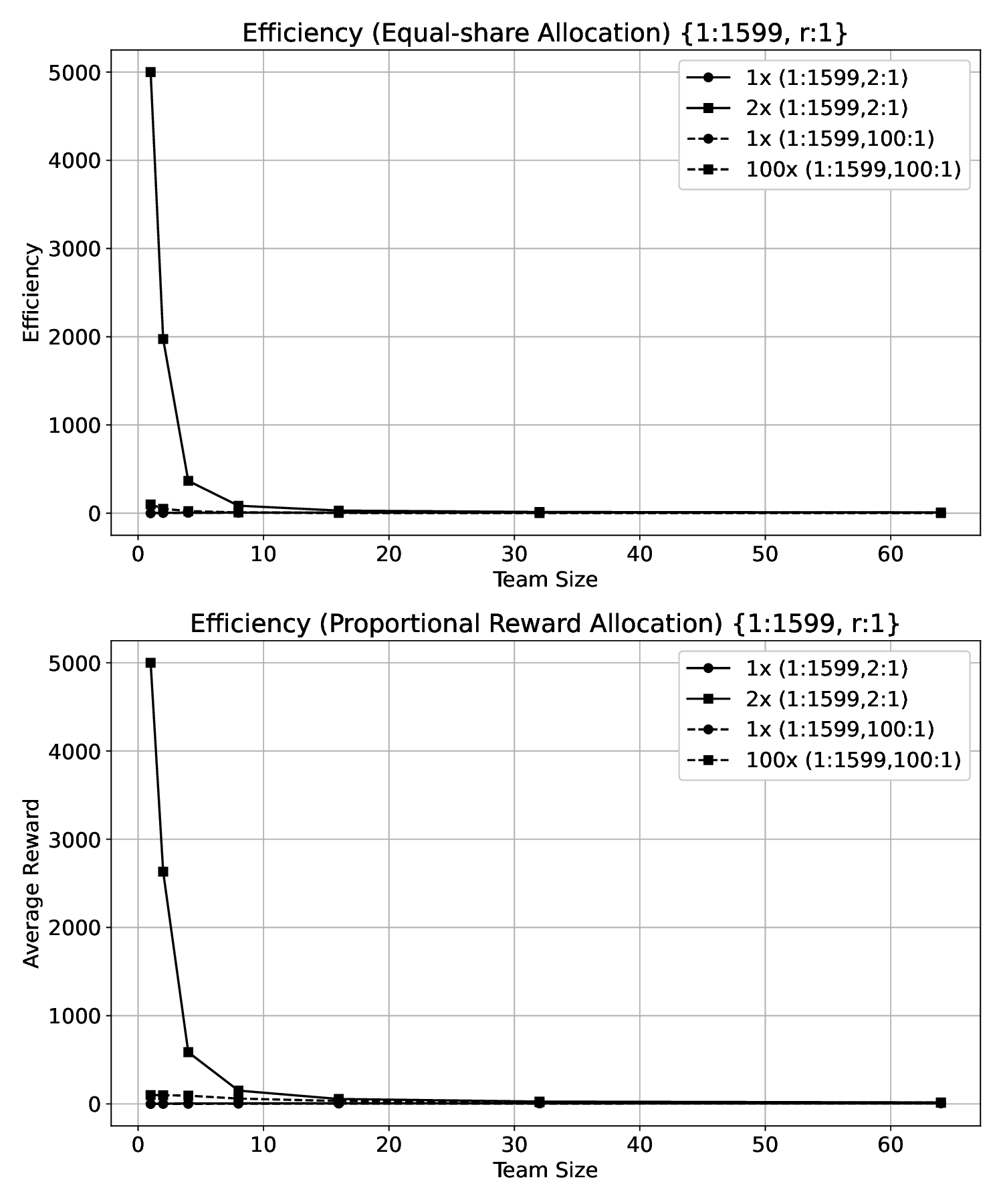}
\caption{Efficiency comparison for {1:1599, $r$:1}.}
\label{fig:efficiency_1599_1}
\end{figure}

For the distribution \{1:1599, $r$:1\}, Figure~\ref{fig:efficiency_1599_1} shows the following trends:

\begin{itemize}
    \item Equal-share allocation: When the performance ratio is $2\times$, smaller teams exhibit high efficiency, but as $M$ increases, efficiency drops rapidly. At $M=64$, efficiency values are 6.24 for low-performance nodes and 8.25 for the high-performance node, meaning no inversion occurs. However, when the performance ratio is $100\times$, at $M=16$, efficiency values are 6.11 for low-performance nodes and 2.37 for the high-performance node, meaning an inversion occurs.
    \item Proportional reward allocation: When the performance ratio is $2\times$ and $M=64$, efficiency values are 6.23 for low-performance nodes and 16.24 for the high-performance node, meaning no inversion occurs. For a $100\times$ performance difference, at $M=64$, efficiency values are 5.41 for low-performance nodes and 13.43 for the high-performance node, again meaning no inversion occurs.
\end{itemize}

\begin{figure}[htbp]
\centering
\includegraphics[width=0.7\linewidth]{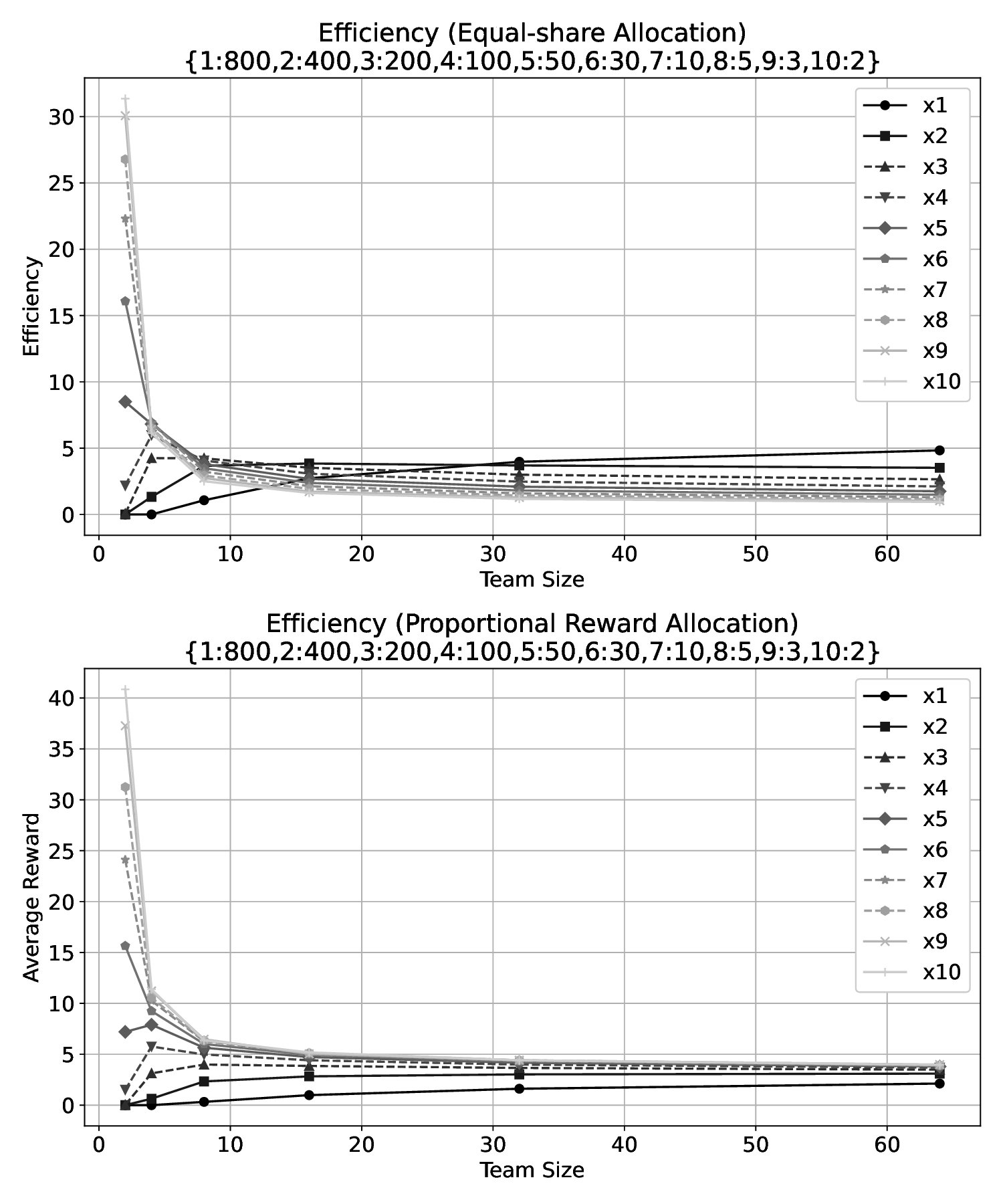}
\caption{Efficiency trends across multiple performance levels.}
\label{fig:efficiency_combined_10layers}
\end{figure}

\paragraph{Multi-Layered Performance Distribution}
A more realistic performance distribution is examined in Figure~\ref{fig:efficiency_combined_10layers}, where multiple performance levels exist: \{1:800,2:400,3:200,4:100,5:50,6:30,7:10,8:5,9:3,10:2\}.

\begin{itemize}
    \item Equal-share allocation: At $M=32$, the lowest-performance (800-node) group achieves the highest efficiency.
    \item Proportional reward allocation: Efficiency ranking remains proportional to node performance, maintaining the expected order across all team sizes.
\end{itemize}

\section{Discussion: Fairness, Economic Incentives, and Sustainability}\label{sec:discussion}
Based on our results, we discuss the implications of PoTS regarding fairness, economic incentives, and long-term sustainability.

\paragraph{Fairness in PoTS}
One of the key takeaways from our analysis is that equal-share allocation fosters greater fairness. Although proportional reward allocation remains attractive for high-performance nodes, equal-share allocation ensures that lower-performance nodes can meaningfully participate. As team sizes increase, low-performance nodes benefit from greater opportunities, making PoTS an inclusive system compared to traditional PoW.

\paragraph{Economic Incentives and Investment Strategies}
For high-performance nodes, proportional reward allocation is clearly preferable as it maximizes their returns. However, our results indicate that, except for extreme cases such as \{1:1599, $r$:1\}, even under equal-share allocation, their rewards do not drastically decrease. This suggests that high-performance nodes still have incentives to participate, but their dominance is reduced compared to PoW.

More importantly, the investment efficiency of high-performance nodes declines in PoTS. The marginal return on computational investment diminishes, meaning that:

\begin{itemize}

\item 
Some participants (``efficiency-focused investors'') may reduce investment in high-performance machines due to lower returns.

\item 
Others (``absolute reward seekers'') may continue participating, but they may reconsider where to allocate their computational resources.
\end{itemize}

This divergence in strategy could lead to reduced overall energy consumption, as efficiency-focused investors are likely to optimize for cost-effectiveness rather than pure computational power.

\paragraph{Implications for Blockchain Sustainability}
A major advantage of PoTS over PoW is its potential to mitigate mining centralization. PoW incentivizes continual investment in high-performance hardware, leading to network centralization. PoTS, by contrast, reduces the financial advantage of high-performance nodes, promoting a more decentralized network.

Additionally, PoTS presents an opportunity to enhance long-term network sustainability by:

\begin{itemize}
\item 
Reducing overall energy consumption, as computational power is not the sole determinant of rewards.

\item 
Encouraging wider participation, ensuring network security through a more diverse set of contributors.

\item 
Offering flexibility in economic models, allowing participants to choose between efficiency-based or absolute-reward-based strategies.
\end{itemize}

Our findings suggest that PoTS is a viable alternative to traditional PoW systems, balancing fairness, investment incentives, and sustainability. While high-performance nodes still receive reasonable rewards, the system prevents excessive centralization and enables broader participation. These characteristics make PoTS an attractive model for future blockchain applications, particularly those prioritizing inclusivity and energy efficiency.

\section{Conclusion and Future Work} \label{sec:conclusion}

Blockchain consensus mechanisms must balance security, decentralization, and fairness to ensure long-term network sustainability. In this paper, we analyzed the fairness properties of Proof of Team Sprint (PoTS), a collaborative consensus algorithm designed to mitigate centralization and enhance participation by lower-performance nodes.

Our study revealed several key findings. First, PoTS significantly improves reward distribution fairness compared to Proof of Work (PoW). Unlike PoW, where rewards are concentrated among high-performance nodes, PoTS ensures that even lower-performance participants receive meaningful incentives, especially as team sizes increase. Second, our simulations demonstrated that reward allocation in PoTS exhibits diminishing returns for high-performance nodes, reducing the incentive for centralization and encouraging a more balanced distribution of computational power. Finally, we found that proportional reward allocation and equal-share allocation influence the degree of fairness differently, with proportional reward allocation maintaining a slight advantage for high-performance participants while still mitigating extreme disparities.

While PoTS presents a promising alternative to traditional mining-based consensus models, several areas for improvement remain. One potential enhancement is refining the reward allocation mechanisms to further optimize fairness and efficiency. For example, hybrid reward structures that dynamically adjust proportional reward allocation and equal-share allocation based on network conditions could enhance system adaptability. Additionally, integrating economic models to predict participant behavior and optimize incentive structures could further improve the overall robustness of PoTS.

Future work should also explore real-world deployment scenarios, including experimental implementations on existing blockchain networks. Evaluating PoTS in practical settings will provide insights into its security resilience, adaptability to network fluctuations, and potential vulnerabilities. Furthermore, integrating PoTS with additional security mechanisms, such as fraud detection and Sybil attack prevention strategies, will be essential for ensuring its viability as a consensus mechanism in diverse blockchain applications.

By fostering a more inclusive and cooperative blockchain environment, PoTS has the potential to reshape the landscape of decentralized consensus. Continued research into its theoretical foundations and real-world applications will be crucial for advancing blockchain fairness, efficiency, and sustainability.

\bibliographystyle{IEEEtran}
\bibliography{fairness_in_pots}

\begin{thebibliography}{10}
\providecommand{\url}[1]{#1}
\csname url@samestyle\endcsname
\providecommand{\newblock}{\relax}
\providecommand{\bibinfo}[2]{#2}
\providecommand{\BIBentrySTDinterwordspacing}{\spaceskip=0pt\relax}
\providecommand{\BIBentryALTinterwordstretchfactor}{4}
\providecommand{\BIBentryALTinterwordspacing}{\spaceskip=\fontdimen2\font plus
\BIBentryALTinterwordstretchfactor\fontdimen3\font minus
  \fontdimen4\font\relax}
\providecommand{\BIBforeignlanguage}[2]{{%
\expandafter\ifx\csname l@#1\endcsname\relax
\typeout{** WARNING: IEEEtran.bst: No hyphenation pattern has been}%
\typeout{** loaded for the language `#1'. Using the pattern for}%
\typeout{** the default language instead.}%
\else
\language=\csname l@#1\endcsname
\fi
#2}}
\providecommand{\BIBdecl}{\relax}
\BIBdecl

\bibitem{nakamoto2008peer}
\BIBentryALTinterwordspacing
S.~Nakamoto, ``Bitcoin: A peer-to-peer electronic cash system,'' 2008.
  [Online]. Available: \url{https://bitcoin. org/bitcoin. pdf}
\BIBentrySTDinterwordspacing

\bibitem{yonezawa2024pots}
\BIBentryALTinterwordspacing
N.~Yonezawa, ``Proof of {Team Sprint}: A collaborative consensus algorithm for
  reducing energy consumption in blockchain systems,'' 2024. [Online].
  Available: \url{https://arxiv.org/abs/2410.12135}
\BIBentrySTDinterwordspacing

\bibitem{afonso2024fairness}
J.~M.~P. Afonso, ``On fairness concerns in the blockchain ecosystem,''
  \emph{arXiv preprint arXiv:2405.18876}, 2024.

\bibitem{huang2021rich}
Y.~Huang, J.~Tang, Q.~Cong, A.~Lim, and J.~Xu, ``Do the rich get richer?
  fairness analysis for blockchain incentives,'' in \emph{Proceedings of the
  2021 international conference on management of data}, 2021, pp. 790--803.

\bibitem{pass2017fruitchains}
R.~Pass and E.~Shi, ``Fruitchains: A fair blockchain,'' in \emph{Proceedings of
  the ACM symposium on principles of distributed computing}, 2017, pp.
  315--324.

\bibitem{amoussou2019fairness}
Y.~Amoussou-Guenou, A.~Del~Pozzo, M.~Potop-Butucaru, and S.~Tucci-Piergiovanni,
  ``On fairness in committee-based blockchains,'' \emph{arXiv preprint
  arXiv:1910.09786}, 2019.

\bibitem{kelkar2020order}
M.~Kelkar, F.~Zhang, S.~Goldfeder, and A.~Juels, ``Order-fairness for byzantine
  consensus,'' in \emph{Advances in Cryptology--CRYPTO 2020: 40th Annual
  International Cryptology Conference, CRYPTO 2020, Santa Barbara, CA, USA,
  August 17--21, 2020, Proceedings, Part III 40}.\hskip 1em plus 0.5em minus
  0.4em\relax Springer, 2020, pp. 451--480.

\bibitem{alamer2024proof}
A.~Alamer and B.~Assiri, ``Proof of fairness: Dynamic and secure consensus
  protocol for blockchain,'' \emph{Electronics}, vol.~13, no.~6, p. 1056, 2024.

\bibitem{zheng2025justitia}
J.~Zheng, H.~Huang, Y.~Liu, T.~Li, H.-N. Dai, and Z.~Zheng, ``Justitia: An
  incentive mechanism towards the fairness of cross-shard transactions,'' in
  \emph{Proc. of IEEE Conference on Computer Communications (INFOCOM)}, 2025,
  pp. 1--10.

\bibitem{orda2021enforcing}
A.~Orda and O.~Rottenstreich, ``Enforcing fairness in blockchain transaction
  ordering,'' \emph{Peer-to-peer Networking and Applications}, vol.~14, no.~6,
  pp. 3660--3673, 2021.

\bibitem{yu2018survey}
Z.~Yu, X.~Liu, and G.~Wang, ``A survey of consensus and incentive mechanism in
  blockchain derived from p2p,'' in \emph{2018 IEEE 24th international
  conference on parallel and distributed systems (ICPADS)}.\hskip 1em plus
  0.5em minus 0.4em\relax IEEE, 2018, pp. 1010--1015.

\bibitem{gunduz2023proof}
F.~G{\"u}nd{\"u}z, S.~Birogul, and U.~Kose, ``Proof of optimum (poo): Consensus
  model based on fairness and efficiency in blockchain,'' \emph{Applied
  Sciences}, vol.~13, no.~18, p. 10149, 2023.

\bibitem{abraham2016solidus}
I.~Abraham, D.~Malkhi, K.~Nayak, L.~Ren, and A.~Spiegelman, ``Solidus: An
  incentive-compatible cryptocurrency based on permissionless byzantine
  consensus,'' \emph{CoRR, abs/1612.02916}, 2016.

\bibitem{yuming2023fairness}
H.~Yuming, ``Fairness and security on blockchain incentives,'' Ph.D.
  dissertation, National University of Singapore (Singapore), 2023.

\bibitem{gudmundsson2024blockchain}
J.~Gudmundsson and J.~L. Hougaard, ``Blockchain-based decentralized reward
  sharing: The case of mining pools,'' \emph{ACM Transactions on Economics and
  Computation}, vol.~12, no.~1, pp. 1--26, 2024.

\bibitem{zhao2022bayesian}
Z.~Zhao, X.~Chen, and Y.~Zhou, ``Bayesian-nash-incentive-compatible mechanism
  for blockchain transaction fee allocation,'' \emph{arXiv preprint
  arXiv:2209.13099}, 2022.

\bibitem{yan2024analyzing}
T.~Yan, S.~Li, B.~Kraner, L.~Zhang, and C.~J. Tessone, ``Analyzing reward
  dynamics and decentralization in ethereum 2.0: An advanced data engineering
  workflow and comprehensive datasets for proof-of-stake incentives,''
  \emph{arXiv preprint arXiv:2402.11170}, 2024.

\bibitem{abraham2023colordag}
I.~Abraham, D.~Dolev, I.~Eyal, and J.~Y. Halpern, ``Colordag: An
  incentive-compatible blockchain,'' \emph{arXiv preprint arXiv:2308.11379},
  2023.

\bibitem{chen2024fairreward}
G.~Chen, C.~Li, W.~Wang, L.~Duan, B.~Wang, Z.~Han, and X.~Zhang, ``Fairreward:
  Towards fair reward distribution using equity theory in blockchain-based
  federated learning,'' \emph{IEEE Transactions on Dependable and Secure
  Computing}, 2024.

\bibitem{li2023reward}
S.-N. Li, F.~Spychiger, and C.~J. Tessone, ``Reward distribution in
  proof-of-stake protocols: A trade-off between inclusion and fairness,''
  \emph{IEEE Access}, vol.~11, pp. 134\,136--134\,145, 2023.

\bibitem{brunjes2020reward}
L.~Br{\"u}njes, A.~Kiayias, E.~Koutsoupias, and A.-P. Stouka, ``Reward sharing
  schemes for stake pools,'' in \emph{2020 IEEE european symposium on security
  and privacy (EuroS\&p)}.\hskip 1em plus 0.5em minus 0.4em\relax IEEE, 2020,
  pp. 256--275.

\bibitem{he2018blockchain}
Y.~He, H.~Li, X.~Cheng, Y.~Liu, C.~Yang, and L.~Sun, ``A blockchain based
  truthful incentive mechanism for distributed p2p applications,'' \emph{IEEE
  access}, vol.~6, pp. 27\,324--27\,335, 2018.

\bibitem{he2022blockchain}
S.~He, Y.~Lu, Q.~Tang, G.~Wang, and C.~Q. Wu, ``Blockchain-based p2p content
  delivery with monetary incentivization and fairness guarantee,'' \emph{IEEE
  Transactions on Parallel and Distributed Systems}, vol.~34, no.~2, pp.
  746--765, 2022.

\bibitem{madupalli2023bimee}
J.~Madupalli and H.~Liu, ``Bimee: Blockchain based incentive mechanism
  considering endowment effect,'' in \emph{2023 Congress in Computer Science,
  Computer Engineering, \& Applied Computing (CSCE)}.\hskip 1em plus 0.5em
  minus 0.4em\relax IEEE, 2023, pp. 1199--1205.

\bibitem{li2023game}
X.~Li, Q.~Liu, S.~Wu, Z.~Cao, and Q.~Bai, ``Game theory based compatible
  incentive mechanism design for non-cryptocurrency blockchain systems,''
  \emph{Journal of Industrial Information Integration}, vol.~31, p. 100426,
  2023.

\bibitem{han2022can}
R.~Han, Z.~Yan, X.~Liang, and L.~T. Yang, ``How can incentive mechanisms and
  blockchain benefit with each other? a survey,'' \emph{ACM Computing Surveys},
  vol.~55, no.~7, pp. 1--38, 2022.

\bibitem{yang2024whales}
W.~Yang, Y.~Kim, T.~H. Kim, C.~H. Lee, and Y.~Ceran, ``From whales to minnows:
  The impact of crypto-reward fairness on user engagement in social media,''
  \emph{Decision Support Systems}, vol. 185, p. 114289, 2024.

\bibitem{pereira2019blockchain}
J.~Pereira, M.~M. Tavalaei, and H.~Ozalp, ``Blockchain-based platforms:
  Decentralized infrastructures and its boundary conditions,''
  \emph{Technological Forecasting and Social Change}, vol. 146, pp. 94--102,
  2019.

\bibitem{xuan2020incentive}
S.~Xuan, L.~Zheng, I.~Chung, W.~Wang, D.~Man, X.~Du, W.~Yang, and M.~Guizani,
  ``An incentive mechanism for data sharing based on blockchain with smart
  contracts,'' \emph{Computers \& Electrical Engineering}, vol.~83, p. 106587,
  2020.

\bibitem{hu2021hybrid}
J.~Hu, M.~J. Reed, M.~Al-Naday, and N.~Thomos, ``Hybrid blockchain for
  iot—energy analysis and reward plan,'' \emph{Sensors}, vol.~21, no.~1, p.
  305, 2021.

\bibitem{mancino2025striking}
D.~Mancino, B.~Guidi, A.~Michienzi, and M.~Viviani, ``Striking the balance:
  Evaluating content quality and reward dynamics in blockchain online social
  media,'' \emph{IEEE Access}, 2025.

\bibitem{xiong2022research}
H.~Xiong, M.~Chen, C.~Wu, Y.~Zhao, and W.~Yi, ``Research on progress of
  blockchain consensus algorithm: A review on recent progress of blockchain
  consensus algorithms,'' \emph{Future Internet}, vol.~14, no.~2, p.~47, 2022.

\bibitem{zhu2020improved}
X.~Zhu, Y.~Li, L.~Fang, and P.~Chen, ``An improved proof-of-trust consensus
  algorithm for credible crowdsourcing blockchain services,'' \emph{IEEE
  access}, vol.~8, pp. 102\,177--102\,187, 2020.

\bibitem{liu2022incentive}
Y.~Liu, Z.~Fang, M.~H. Cheung, W.~Cai, and J.~Huang, ``An incentive mechanism
  for sustainable blockchain storage,'' \emph{IEEE/ACM Transactions on
  Networking}, vol.~30, no.~5, pp. 2131--2144, 2022.

\end{thebibliography}

\end{document}